\documentclass[reprint,aps,pra,groupedaddress,showpacs]{revtex4-1}
\usepackage{bm}
\usepackage{graphicx}
\usepackage{color}
\usepackage{amsmath}
\usepackage{natbib}
\usepackage{url}
\usepackage{textcase}
\usepackage{amsthm}
\usepackage{mathcomp}
\usepackage{amsfonts}
\usepackage{amssymb}
\usepackage{mathtools}

\begin{document}

\title{Nonlinear reversal of PT-symmetric phase transition in a system of coupled semiconductor micro-ring resonators}

\author{Absar U. Hassan}
\email{absar.hassan@knights.ucf.edu}
\author{Hossein Hodaei}
\author{Mohammad-Ali Miri}
\author{Mercedeh Khajavikhan}
\author{Demetrios N. Christodoulides}

\affiliation{CREOL/College of Optics and Photonics, University of Central Florida, Orlando, Florida 32816, USA}

\date{\today}

\begin{abstract}
A system of two coupled semiconductor-based resonators is studied when lasing around an exceptional point. We show that the presence of nonlinear saturation effects can have important ramifications on the transition behavior of this system. In sharp contrast with linear PT-symmetric configurations, nonlinear processes are capable of reversing the order in which the symmetry breaking occurs. Yet, even in the nonlinear regime, the resulting non-Hermitian states still retain the structural form of the corresponding linear eigenvectors expected above and below the phase transition point. The conclusions of our analysis are in agreement with experimental data.
\end{abstract}


\maketitle

\section{\label{S1}Introduction}
In recent years there has been a growing interest in optical structures based on parity-time (PT) symmetry ~\cite{Makris,Ramy,Guo,Klaiman,LonghiPRA,Ruter,Regensburger,Eva2011,Kottos,Sukov}. Along these lines, some intriguing possibilities have been predicted and experimentally demonstrated. These include solitons~\cite{Ziad,MiriSoliton}, Bloch oscillations and exceptional lines in PT-symmetric lattices~\cite{LonghiPRL,Makris3,Soljacic}, unidirectional invisibility~\cite{Unidirectional,Regensburger,Feng3}, power oscillations~\cite{Ruter,Regensburger,Makris2}, and mode management in laser structures~\cite{MiriOL,Hodaei,PTcavities,*PToptica}, to mention a few. In optical realizations, PT-symmetry can be established by judiciously incorporating gain and loss in a given structure. In particular, a necessary condition for this symmetry to hold is that the complex refractive index distribution involved should obey  $n(\bm{r})=n^*(-\bm{r})$. In other words, the real part of the refractive index profile must be an even function of position whereas its imaginary counterpart must be odd~\cite{Makris}. Under these latter conditions, an optical system (composed of cavities, waveguides etc), can behave in a pseudo-Hermitian fashion provided that the overall attenuation and amplification is appropriately balanced. On the other hand, if the gain/loss contrast exceeds a certain threshold, the PT-symmetry can be spontaneously broken and the spectrum is no longer entirely real~\cite{Bender1998,Znojil,Ahmed}. This marks the presence of an exceptional point~\cite{Kato,Berry,WDHeiss,Ramy2} or the emergence of a PT-symmetry breaking transition~\cite{MXiaoPTcoupled,BPengPTcoupled,Stone}. Moreover, a number of studies have suggested that PT-symmetric concepts can also be fruitfully utilized in other settings beyond optics~\cite{TKottosElectronicPT,AluAcoustics,Kottos2,Kottos3}.

Lately, the phase transitions associated with exceptional points, have been effectively utilized to enforce single-mode operation in micro-ring laser resonators~\cite{Hodaei,FengPTlaser} and pump-induced lasing turn-off~\cite{SRotterPumpInducedEPs,SRotterReversing}. In particular, in Ref.~\cite{Hodaei}, it was shown that the selective breaking of PT-symmetry can be exploited to enhance the maximum output power in a desired longitudinal mode. This mode selection scheme is inherently self-adapting and can be used over a broad bandwidth without the need of any other intra-cavity components. While the mechanism of PT-symmetry breaking is inherently linear, of fundamental interest will be to understand how this process unfolds in the presence of nonlinearity. This is imperative given that lasers are by nature nonlinear devices. In Ref.~\cite{SegevNonlinearPT} Lumer et al. already indicated that it is possible to reverse the PT phase transition sequence using the conservative component of a Kerr nonlinearity in a periodic structure. Instead, in this paper we consider the properties of PT-symmetric coupled micro-ring laser cavities in the presence of saturation effects associated with the imaginary part of the nonlinearity.

In what follows we provide a nonlinear model describing the field evolution in two coupled cavities in the presence of saturable gain and loss that is prevalent in semiconductor systems. A dual micro-ring arrangement is studied using a temporal coupled mode formalism~\cite{HausTemporal} when one ring is subjected to optical pumping while its counterpart is kept un-pumped. The system is shown to move from the linear broken PT-symmetry domain directly into the nonlinear broken regime. It is further demonstrated that by increasing the pumping level, the eigenmodes of the system transition into an unbroken pair of PT-symmetric states that exhibit two real eigenvalues. Conversely, if the system starts lasing in an unbroken PT-like phase then it remains there in spite of nonlinear saturation effects. An experimental demonstration of this process, when starting from the broken phase, is presented which is in qualitative agreement with our theoretical results. Finally, we briefly discuss the behavior of this nonlinear system starting from having both micro-rings equally pumped to eventually blocking the pump from one of them.
\section{\label{S2}Theoretical Model}
\begin{figure}
  \includegraphics{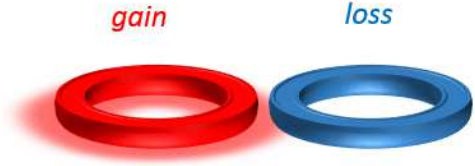}
  \caption{\label{fig1}A PT-symmetric arrangement of two coupled micro-ring resonators.}
\end{figure}

A schematic of a dual micro-ring arrangement is shown in Fig.~\ref{fig1}. Each micro-ring in our system involves a multiple quantum well InGaAsP-InP structure that is embedded in a silica substrate as in Ref.~\cite{Hodaei}. The top surface of the rings is exposed to air that serves as a cladding. At the operating wavelength of $1.55$ $\mu$m of this quantum well system, the effective refractive index is $n_e\simeq3$ while the group index in the waveguide rings is $n_g\simeq4$.
For demonstration purposes here we assume that each cavity supports only a single longitudinal and transverse mode. In general, the dynamics in each cavity in isolation are described by a corresponding set of modal field amplitude equations in conjunction with a carrier evolution equation. Yet, once the carriers attain a steady-state, the field equation can be further simplified according to~\cite{Agrawal}:
\begin{equation}
\label{model}
\frac{dE}{dt}=\frac{1}{2}\left(\sigma\frac{(p-1)}{1+\varepsilon|E|^2}-\gamma_p\right)\left(1-i\alpha_H\right)E
\end{equation}
Here $p=\tau_e R_p/N_0$ is a pump parameter, where the carrier generation rate, $R_p=\eta I/(\hbar\omega d)$ and $I$,$\eta$, $d$ are the pump intensity, the external quantum efficiency and the depth of each micro-ring respectively. In addition, $\tau_e$ represents the carrier lifetime, $N_0$ stands for the transparency carrier population density, $\omega$ is the frequency of the emitted light and $\hbar$ is the Planck\textquoteright s constant.  The parameter $\varepsilon$ is inversely proportional to the saturation intensity $\varepsilon=n_e c\epsilon_0\Gamma a\tau_e/(2\hbar\omega)$ and $\sigma$ is proportional to the saturated loss in the absence of pumping $(\sigma=\Gamma v_g a N_0)$. The linear loss $\gamma_p$ is the inverse of the photon lifetime $(\tau_p)$ in the cavity and $\alpha_H$ is the linewidth enhancement factor. Finally, $c \text{ and } \epsilon_0$ are the speed of light and permittivity in vacuum respectively, $\Gamma$ is the confinement factor, $v_g$ represents the group velocity and $a$, the gain constant ($g=a(N-N_0)$).

Note that in formulating the evolution equations the material response is assumed to be fast compared to carrier and photon lifetimes and hence is considered here to be instantaneous~\cite{Agrawal}. In our arrangement we assume that the coupling strength between the two rings is strong and hence any frequency detuning that could result from the $\alpha_H$-factor can be ignored. By adopting this latter assumption and denoting the unsaturated gain as $\tilde{g_0}=\sigma(p-1)$, unsaturated loss (at $p=0$) as $\tilde{f_0}=\sigma$, we obtain the following two equations describing the dynamics in the aforementioned coupled cavities,
\begin{subequations}
 \label{pre full system}
 \begin{eqnarray}\
 \frac{dA_1}{dt}=-\tilde{\gamma} A_1+\left(\frac{\tilde{g_0}}{1+\varepsilon|A_1|^2}\right)A_1+i\kappa A_2
 \\
 \frac{dA_2}{dt}=-\tilde{\gamma} A_2-\left(\frac{\tilde{f_0}}{1+\varepsilon|A_2|^2}\right)A_2+i\kappa A_1.
 \end{eqnarray}
\end{subequations}
In Eq.~(\ref{pre full system}), the modal field amplitudes $A_1$,$A_2$ correspond to the pumped and un-pumped resonators respectively, $\tilde{\gamma}$ is the linear loss present in both cavities and $\kappa$ is the coupling strength between the resonators.
A normalized version of these equations can be easily obtained by adopting the normalized quantities, $a_{1,2}=\sqrt{\varepsilon}A_{1,2}$, $\tau=\kappa t$, $\gamma = \tilde{\gamma}/\kappa$, $f_0 = \tilde{f_0}/\kappa$ and $g_0 = \tilde{g_0}/\kappa$,
\begin{subequations}
 \label{full system}
 \begin{eqnarray}
 \dot{a_1}=-\gamma a_1+\left(\frac{g_0}{1+|a_1|^2}\right)a_1+i a_2\label{coupled 1}
 \\
 \dot{a_2}=-\gamma a_2-\left(\frac{f_0}{1+|a_2|^2}\right)a_2+i a_1\label{coupled 2}
 \end{eqnarray}
\end{subequations}
where $\dot{a}=da/d\tau$. In what follows we will study the behavior associated with this system of nonlinear evolution equations.
\section{\label{S3}Linear dynamical analysis}
To analyze the response of this arrangement under linear conditions, we assume that the modal field amplitudes are small, i.e. $|a_{1,2}|\sim0$. Under these assumptions, saturation effects in both the gain and loss mechanisms can be ignored. Hence, this regime can be effectively described by a linearized version of Eq.~(\ref{full system}), e.g.,
\begin{subequations}
\label{linear system}
\begin{eqnarray}
\dot{a_1}=-\gamma a_1 + g_0 a_1 + i a_2\label{linear 1}
\\
\dot{a_2}=-\gamma a_2 - f_0 a_2 + i a_1\label{linear 2}
\end{eqnarray}
\end{subequations}
The eigenvalues of this system, $\lambda$, can be directly obtained by adopting the form, $\begin{pmatrix} a_{1}&a_{2} \end{pmatrix}^{T}= \begin{pmatrix} a_{01}&a_{02} \end{pmatrix}^{T} e^{-i\lambda \tau}$, where $a_{01,02}$ are complex constants and $T$ represents a transpose operation. In this respect, two regimes can be identified depending on whether $(g_{0}+f_{0}) \lessgtr 2$. In the first case where $(g_0 + f_0 )<2 $, the modal solutions of Eq.~(\ref{linear system}) are given by,
\begin{equation}
\label{linear unbroken}
\begin{pmatrix} a_1\\a_2 \end{pmatrix}=\begin{pmatrix} 1\\\pm e^{\pm i\theta} \end{pmatrix}e^{(\frac{g_0-f_0}{2}-\gamma )\tau}e^{\pm i( \cos \theta) \tau},
\end{equation}
where $\sin ⁡\theta =(g_0+f_0)/2$. We note that the structure of the modal fields closely resembles that expected from an unbroken PT-symmetric coupled arrangement~\cite{MiriOL}. In particular, the two eigenvectors are by nature non-orthogonal with a phase factor $\theta$ that depends on the gain/loss contrast. In addition, a PT-like bifurcation is present around a threshold value given that,
\begin{figure*}
  \includegraphics{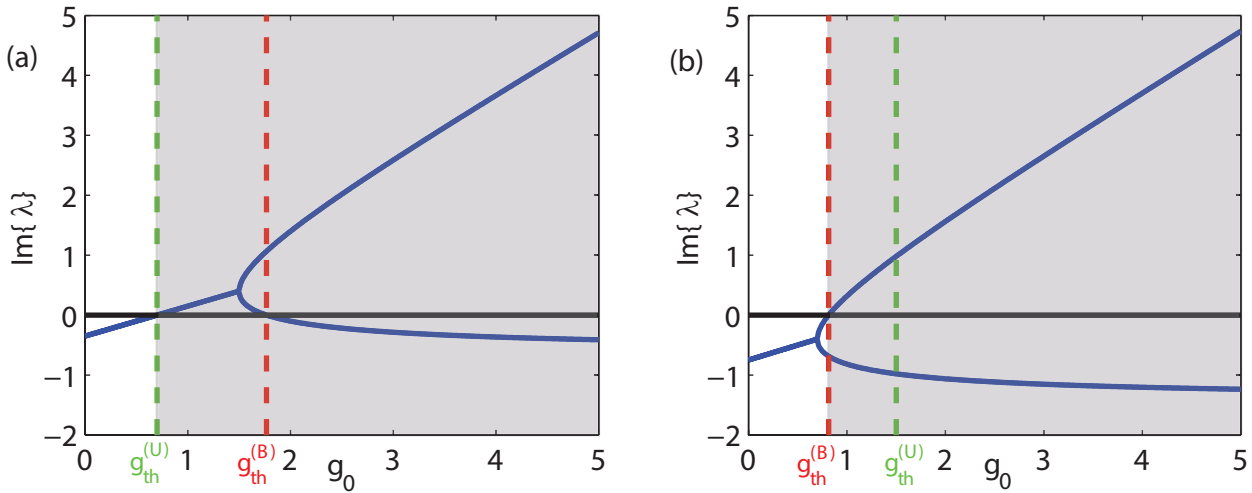}
  \caption{\label{fig2}Imaginary components of eigenvalues (blue curves) of the linear system are displayed as the gain level increases. In all cases amplification occurs if $\textit{Im}\left\{\lambda\right\}>0$\textemdash represented by the gray regions. The broken PT-symmetric phase appears after a bifurcation takes place. The graph in (a) shows that $\textit{Im}\left\{\lambda \right\}>0$ before branching occurs, i.e. when $(\gamma+f_0)<1$, whereas in (b) lasing begins in the broken phase, which takes place when $(\gamma+f_0)>1$. In both cases the dashed lines indicate the two possible thresholds, where the red line corresponds to the broken phase ($g_{\text{th}}^{(\text{B})}=\gamma+1/(\gamma+f_0)$) and the green to the unbroken ($g_{\text{th}}^{(\text{U})}=2\gamma+f_0$). The system parameters used here are $\gamma=0.1$ and (a) $f_0=0.5$, (b) $f_0=1.3$.}
\end{figure*}
\begin{equation}
\label{linear bifurcation}
\cos \theta = \sqrt{1-\left(\frac{g_0+f_0}{2}\right)^2}
\end{equation}
If on the other hand $(g_0+f_0 )>2$, the eigenvectors of Eq.~(\ref{linear system}) are,
\begin{equation}
\label{linear broken}
\begin{pmatrix} a_1\\a_2 \end{pmatrix}=\begin{pmatrix} 1\\i e^{\pm \theta} \end{pmatrix}e^{(\frac{g_0-f_0}{2}-\gamma )\tau}e^{\mp ( \sinh \theta) \tau},
\end{equation}
where $\cosh⁡ \theta = (g_0+f_0)/2$. As opposed to those described by Eq.~(\ref{linear unbroken}), these latter solutions exhibit features of a broken PT-symmetric configuration. In this regime the modal field amplitudes are phase shifted by $\pi/2$ and moreover, they are unequal.

If the system is operating in the first regime (unbroken PT-symmetry, given by Eq.~(\ref{linear unbroken}), then the fields will experience linear amplification as long as the gain is above the total loss in the system, i.e. $g_0>(2\gamma+f_0)= g_{\text{th}}^{\text{(U)}}$. Conversely, in the broken PT-symmetric phase (described by Eq.~(\ref{linear broken})), growth will occur provided that,
\begin{equation}
\label{ineq}
\left(\frac{g_0-f_0}{2}-\gamma\right)+\sqrt{\left(\frac{g_0+f_0}{2}\right)^2-1} > 0.
\end{equation}
Equation ~(\ref{ineq}) implies that in this case, the threshold for lasing is dictated by the following condition,
\begin{equation}
\label{linear broken gain range}
g_0 > \frac{1}{\gamma+f_0}+\gamma,
\end{equation}
i.e. the gain threshold in this broken symmetry is $g_{\text{th}}^{\text{(B)}}=(\gamma+f_0)^{-1}+\gamma$. In view of the above results, one can conclude that the lasing thresholds corresponding to these two regimes (above/below the PT-symmetry breaking point) are uniquely determined by the parameters, $\gamma$, and  $f_0$. To compare these thresholds, one has to consider whether $(\gamma+f_0)\gtrless 1$. If for example, $(\gamma+f_0)> 1$ then the broken phase (Eq.~(\ref{linear broken})) has a lower threshold and therefore will lase ($g_{\text{th}}^{\text{(B)}}<g_{\text{th}}^{\text{(U)}}$). Interestingly however, if $(\gamma+f_0)< 1$, the situation is reversed and the unbroken PT eigenstate, as given by Eq.~(\ref{linear unbroken}), will experience amplification.
The behavior of the system in these two different domains is depicted in Figs.~\ref{fig2}(a) and ~\ref{fig2}(b) for various values of the gain, $g_0$. Figure~\ref{fig2}(b) clearly suggests that for $(\gamma+f_0)>1$ $(\text{i.e. } g_{\text{th}}^{\text{(B)}}<g_{\text{th}}^{\text{(U)}})$, the lasing threshold is in fact lower than the total loss in the system, $(f_0+2\gamma)$. This counter-intuitive result is attributed to the coupling process which is in this case relatively slow and therefore does not allow the photon energy to see the entire two-ring system. These two lasing thresholds can be summarized by the following inequality,
\begin{equation}
\label{global gain threshold}
g_0 > \min\left[\frac{1}{\gamma+f_0}+\gamma, 2\gamma+f_0\right].
\end{equation}
With these preliminary conditions for $g_0$, needed for lasing, we can now consider the ensuing nonlinear response of this system.
\section{\label{S4}Nonlinear regimes}
As the fields in the PT-coupled cavity configuration start to grow, nonlinear saturation effects come into play, as described by Eq.~(\ref{full system}). Yet, as we will see, the properties of the linear system not only determine the lasing thresholds, but also provide valuable information as to how this arrangement will respond in the nonlinear regime. More specifically, if $(\gamma + f_0)>1$, the system will start from a linear broken PT-symmetry and then enter a broken PT-like nonlinear domain. By further increasing the gain, this same arrangement will transition into a nonlinear unbroken PT phase and will remain there. If on the other hand, $(\gamma + f_0)<1$, this structure will lase into an unbroken PT-like domain (whether linear or nonlinear) for all values of the gain $g_0$ above threshold. It is important to emphasize that in the first case of $(\gamma + f_0)>1$, upon increasing the pump level, a reversal in the order in which symmetry breaking occurs is observed, i.e. the solutions transition from a broken to an unbroken state. The two possible nonlinear phases of lasing are described below along with their corresponding gain parameter ranges.
\subsection{\label{S4-1}Broken PT}
This section is pertinent to the case where lasing takes place in the linear PT-broken regime where $(\gamma+f_0)>1$. In this scenario, the nonlinear broken-PT supermodes can be directly obtained by assuming stationary solutions for the field amplitudes that have the form, $\begin{pmatrix} a_{1}&a_{2} \end{pmatrix}^{T}= \begin{pmatrix} a_{01}&a_{02} \end{pmatrix}^{T}$, where $a_{01,02}$ are complex constants. Equation~(\ref{full system}) is then reduced to,
\begin{subequations}
\label{nonlinear broken coupled}
\begin{eqnarray}
0=-\gamma a_{01} + \left(\frac{g_0}{1+|a_{01}|^2}\right)a_{01}+i a_{02}\label{nonlinear broken 1}
\\
0=-\gamma a_{02} - \left(\frac{f_0}{1+|a_{02}|^2}\right)a_{02}+i a_{01}\label{nonlinear broken 2}
\end{eqnarray}
\end{subequations}
These equations clearly suggest that $a_{01}$ and $a_{02}$ are out of phase by $\pi/2$. This in turn allows one to write $a_{02} = i \rho a_{01}$, where $\rho \in \Re^{+}$ (see Eq.~(\ref{nonlinear broken 2})) represents the modal ratio. From Eq.~(\ref{nonlinear broken coupled}), we readily obtain the following quartic polynomial equation for $\rho$,
\begin{eqnarray}
\rho^4 - \left(g_0 +\frac{1}{\gamma}-\gamma\right)\rho^3+\left(\frac{g_0-f_0}{\gamma}-2\right)\rho^2\label{quartic broken}\nonumber\\+\left(-f_0+\frac{1}{\gamma}-\gamma\right)\rho+1=0
\end{eqnarray}

In solving Eq.~(\ref{quartic broken}), we look for a real root in the interval $\left[0, 1\right]$ since, from a physical perspective, one expects that under steady state conditions, the modal field in the lossy ring will be less than that with gain. In addition, one can show that among all four possible roots, that contained in $\left[0, 1\right]$ happens to be the only stable one. It is important to note that similar to the broken symmetry modes in linear PT systems, the solution sets in this regime are characterized by an asymmetric distribution of modal fields in the two coupled resonators. For this specific reason, the point $\rho=1$ is crucial since it marks a PT-breaking transition. The critical gain value ($g_c$) where this transition occurs is found to be,
\begin{equation}
g_c=f_0\frac{(1+\gamma)}{(1-\gamma)}.
\end{equation}
\begin{figure}
  \includegraphics{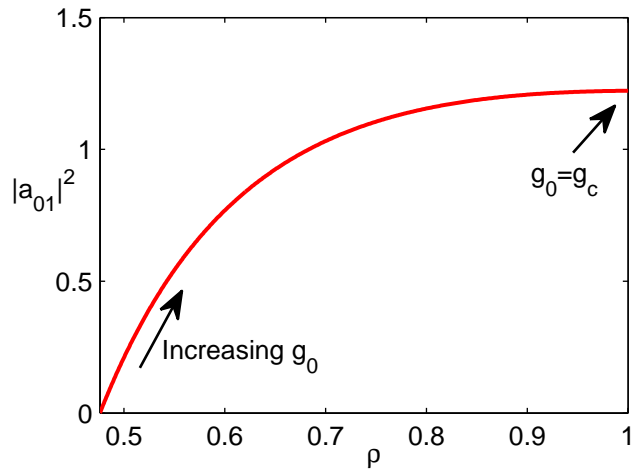}
  \centering
  \caption{\label{fig3}Light intensity in the pumped ring as a function of the modal ratio $\rho$, as obtained from Eqs.~(\ref{nonlinear broken coupled}) and~(\ref{quartic broken}) when $f_0=2$ and $\gamma=0.1$. The linear gain $g_0$ is varied between $g_{\text{th}}^{\text{(B)}}=0.58 \text{ to } g_c = 2.44$}
\end{figure}
Note that this critical gain value is smaller than the lasing threshold needed for the linear unbroken phase, $g_{\text{th}}^{\text{(U)}}$, that is possible in the parameter range $(\gamma+f_0)<1$. Hence this nonlinear broken PT phase only arises once lasing begins in the linear broken PT phase which only occurs when $(\gamma+f_0)>1$, where we also have $g_c>g_{\text{th}}^{\text{(B)}}$. To demonstrate the energy occupancy in the two cavities, we vary the value of $g_0$ in the range $g_{\text{th}}^{\text{(B)}}<g_0<g_c$. Figure.~\ref{fig3} depicts these results for $\gamma=0.1$ and $f_0=2$.

As it can be seen in Fig.~\ref{fig3}, higher values of $g_0$ not only result in higher steady state intensities in the resonators but also lead to an increased ratio ($\rho$) that eventually becomes unity. As previously mentioned, an unequal distribution of the fields in the two rings, along with a phase difference of $\pi/2$ clearly indicates that the solution sets in this regime have broken PT-like forms as in Eq.~(\ref{linear broken}). Moreover, there is no frequency shift associated with the resonance of the ring system\textemdash another indicator of a broken PT-symmetry.
\begin{figure}
  \includegraphics{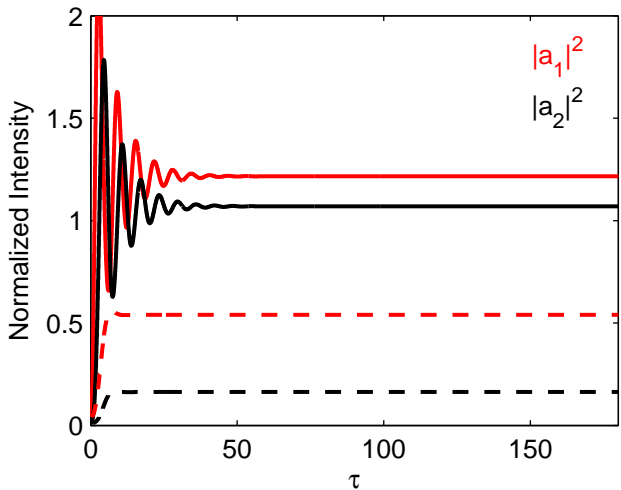}
  \centering
  \caption{\label{fig4}The unequal distribution of steady state intensities (broken symmetry) in the rings with gain (red) and loss (black) is shown. The curves are obtained after numerically integrating Eq.~(\ref{full system}) for $\gamma=0.1$ and $f_0=2$. Dashed lines represent the solution for $g_0=1$ and solid lines are obtained for $g_0=2.3$. A higher gain naturally results in higher intensities but at the same time, the intensity contrast between the two resonators decreases.}
\end{figure}

Remarkably, if one considers linear PT-symmetric dimers, it is well known that the difference between the field intensities in the two components of the dimer becomes larger as we increase the gain-loss contrast beyond the spontaneous symmetry breaking point, indicated by the term $e^{\pm\theta}$ in Eq.~(\ref{linear broken}).
However, in the nonlinear case, higher pumping levels (larger values of gain) eventually lead the system to the symmetric phase, at $\rho=1$. Figure~\ref{fig4} shows the time evolution of the modal intensities as a function of $\tau$ after solving Eq.~(\ref{full system}) for $g_0=1$ and $g_0=2.3$. This latter figure demonstrates that higher pump levels eventually enforce a transition towards an unbroken phase where the modal ratio is unity. To summarize, the relevant gain range for solutions within this regime is $g_{\text{th}}^{\text{(B)}}<g_0<g_c$, provided that $(\gamma+f_0)>1$.
\subsection{\label{S4-2}Unbroken PT}
Before we discuss in detail the properties associated with the nonlinear unbroken PT-symmetry, we note that the results of this section are applicable in both regimes, i.e. $(\gamma+f_0)\gtrless1$. To obtain the nonlinear eigenmodes in the PT-symmetric phase, we now assume time harmonic solutions, $\begin{pmatrix} a_{1}&a_{2} \end{pmatrix}^{T}= \begin{pmatrix} a_{01}&a_{02} \end{pmatrix}^{T}e^{i\lambda \tau}$, where $\lambda\in\Re$.
In this case, Eq.~(\ref{full system}) leads to the following relations:
\begin{subequations}
\label{nonlinear unbroken coupled}
\begin{eqnarray}
i\lambda a_{01}=-\gamma a_{01} + \left(\frac{g_0}{1+|a_{01}|^2}\right)a_{01}+i a_{02}\label{nonlinear unbroken 1}
\\
i\lambda a_{02}=-\gamma a_{02} - \left(\frac{f_0}{1+|a_{02}|^2}\right)a_{02}+i a_{01}\label{nonlinear unbroken 2}.
\end{eqnarray}
\end{subequations}
Using the representation $g_s=g_0/(1+|a_{01}|^2)$ for the saturated gain and $f_s=f_0/(1+|a_{02}|^2)$ for the saturated loss and assuming that $a_{01,02}\neq 0$, we get a quadratic equation for the eigenvalues,
\begin{equation}
\label{unbroken eigenvalue equation}
\lambda^2-i(2\gamma+f_s-g_s)\lambda-(\gamma^2+\gamma(f_s-g_s)-g_sf_s+1)=0.
\end{equation}
Given that $\lambda$ is real, it is necessary that,
\begin{equation}
\label{leading to cosh etc}
\left(\frac{g_s}{2\gamma}\right)-\left(\frac{f_s}{2\gamma}\right)=1.
\end{equation}
 This last relation is directly satisfied through the parametric representation $g_s=2\gamma \cosh^2⁡(\eta)$ and $f_s=2\gamma \sinh^2(\eta)⁡$ where $\eta$ is a positive real quantity. In this respect we arrive at the following relations for the intensities,
\begin{eqnarray}
\label{a01-cosh}
|a_{01}|^2=\frac{g_0}{2\gamma\cosh^2(\eta)}-1
\\
\label{a02-sinh} |a_{02}|^2=\frac{f_0}{2\gamma\sinh^2(\eta)}-1.
\end{eqnarray}

The eigenvalue Eq.~(\ref{unbroken eigenvalue equation}), now readily reduces to, $\lambda^2=1-\gamma^2\cosh^2(2\eta)$, in which case, $\lambda_{1,2}=\pm\cos\theta_{\text{nl}}$ provided that $\gamma\cosh⁡(2\eta)=\sin\theta_{\text{nl}}$. Here $\theta_{\text{nl}}$ represents a nonlinear phase shift ranging between $0$ and $\pi/2$. Moreover, after dividing Eq.~(\ref{nonlinear unbroken 1}) by $a_{01}$ and Eq.~(\ref{nonlinear unbroken 2}) by $a_{02}$, upon subtraction we obtain,
\begin{multline}
\label{phi,rho}
2i\sin\theta_{\text{nl}}=(\rho-\rho^{-1})\cos\phi+i(\rho+\rho^{-1})\sin\phi
\end{multline}
\begin{figure}
  \includegraphics{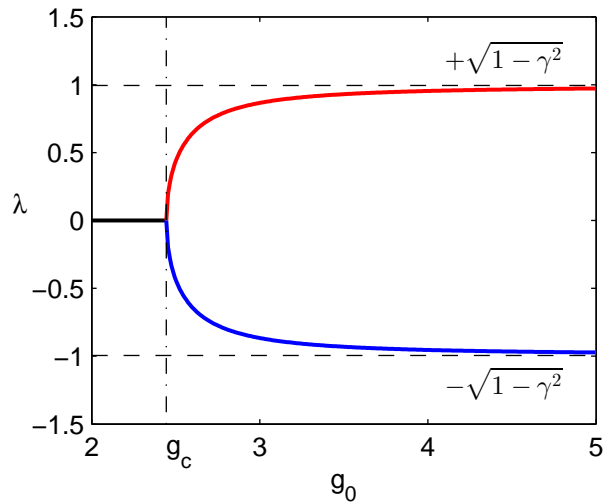}
  \centering
  \caption{\label{fig5}The eigenvalues of the nonlinear system exhibit a square-root bifurcation when entering the unbroken symmetry regime. The region $g_0<g_c$ represents broken symmetry where the eigenvalues are degenerate. The parameter values used here are $\gamma=0.1$ and $f_0=2$. The eigenvalues $\lambda_{\pm}$ approach the asymptotes $\pm \sqrt{(1-\gamma^2)}$ for large values of $g_0$.}
\end{figure}
where $a_{02}=\rho e^{i\phi}a_{01}$. Equation~(\ref{phi,rho}) can be solved for the real and imaginary parts, from where one finds that, $\rho=\pm 1$ and $\phi = \pm \theta_{\text{nl}}$, which clearly suggests that $|a_{01}|^2=|a_{02}|^2$. This is only possible as long as (by considering Eqs.~(\ref{a01-cosh}) and (\ref{a02-sinh})),
\begin{equation}
\label{tanh}
\tanh(\eta)=\sqrt{\frac{f_0}{g_0}}
\end{equation}
i.e. $g_0>f_0$. The eigenvalue expression in Eq.~(\ref{unbroken eigenvalue equation}) now simplifies to,
\begin{equation}
\label{nonlinear eigenvalues}
\lambda^2 = 1 - \gamma^2\left(\frac{g_0+f_0}{g_0-f_0}\right)^2.
\end{equation}
Equation~(\ref{nonlinear eigenvalues}) directly indicates that real eigenvalues are only possible if, $g_0\geq f_0(1+\gamma)/(1-\gamma)$, which is equivalent to $g_0\geq g_c$, corroborating the earlier findings in Sec.~\ref{S4-1}. In other words the gain level has to be above this critical value, a necessary condition for observing solution sets in this regime. The unfolding of the nonlinear eigenvalues as a function of the gain level is shown in Fig.~\ref{fig5}.
From these results, one can then determine the unbroken nonlinear PT-symmetric eigenvectors, e.g.,
\begin{equation}
\label{nonlinear unbroken solution}
\begin{pmatrix} a_1\\a_2 \end{pmatrix}=\sqrt{\left(\frac{g_0-f_0}{2\gamma}-1\right)}
\begin{pmatrix} 1\\ \pm e^{\pm i\theta_{\text{nl}}}\end{pmatrix}e^{\pm i(\cos\theta_{\text{nl}})\tau},
\end{equation}
where $\sin\theta_{\text{nl}}=\gamma(g_0+f_0)/(g_0-f_0)$. When Equations~(\ref{a01-cosh}) and (\ref{a02-sinh}) are used in conjunction with Eq.~(\ref{tanh}), they provide another restriction on the value of $g_0$ since $|a_i|^2>0$. More specifically, the restriction is given by $g_0\geq(2\gamma+f_0)=g_{\text{th}}^{\text{(U)}}$. Hence, the complete range of $g_0$ for this solution to exist is:
\begin{equation}
\label{gain range NL unbroken}
g_0 \geq g_c \cap  g_0\geq g_{\text{th}}^{\text{(U)}}
\end{equation}
\begin{figure}
  \includegraphics{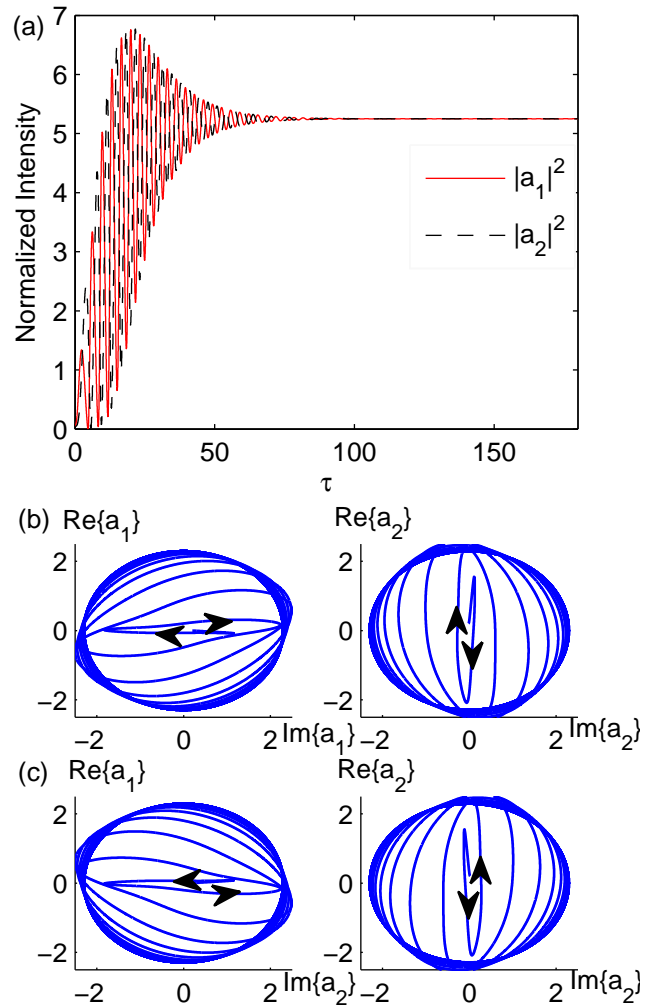}
  \centering
  \caption{\label{fig6}(a)  Intensity evolution in the two rings is plotted against time $\tau$, when, $g_0=2.25$, $f_0=1$ and $\gamma=0.1$. Trajectory of the modal fields when (b)$\Delta\phi=\pi/2+0.1$\textemdash clockwise rotation and (c)$\Delta\phi=\pi/2-0.1$\textemdash counter-clockwise rotation.}
\end{figure}
It should be noted here that under the condition $(\gamma+f_0)>1$, i.e. when lasing begins in the broken PT phase, once the gain level exceeds $g_c$, both conditions in (\ref{gain range NL unbroken}) are satisfied and the steady state now assumes the nonlinear unbroken form of Eq.~(\ref{nonlinear unbroken solution}). This confirms the aforementioned reversal of PT-symmetric phase transition due to the nonlinearity. However, if $(\gamma+f_0)<1$, where lasing begins in the linear unbroken PT phase, the lasing threshold $g_{\text{th}}^{\text{(U)}}$ is greater than $g_c$, which immediately implies that once lasing begins, the system will eventually attain the nonlinear unbroken PT-symmetric steady state solutions, described by Eq.~(\ref{nonlinear unbroken solution}).

The time evolution of intensities in the two coupled resonators can be studied by numerically solving Eq.~(\ref{full system}). These results are displayed in Fig.~\ref{fig6}(a). Notice that in this domain, the modal amplitudes eventually become equal (irrespective of initial conditions), an indication of a nonlinear unbroken PT-symmetry. The analytical expressions in Eq.~(\ref{nonlinear unbroken solution}) suggest that the system admits two fixed point solutions, $\lambda_{1,2}=\pm\cos\theta_{\text{nl}}$. The choice between the two depends upon the initial conditions provided. An effective method of deducing which initial conditions correspond to which of the two attractors, is to project the initial vector onto the two nonlinear eigenmodes.

The projection operation has to be carried out in a PT-symmetric sense~\cite{BenderQM}, i.e. respecting bi-orthogonality, which implies that for two vectors $\Phi_1=\begin{pmatrix} \phi_{1x}&\phi_{1y}\end{pmatrix}^{T}$ and $\Phi_2=\begin{pmatrix} \phi_{2x}&\phi_{2y}\end{pmatrix}^{T}$,
\begin{equation}
\label{PT projection}
\langle \Phi_1 | \Phi_2\rangle = \begin{pmatrix} \phi_{1y}^*&\phi_{1x}^*\end{pmatrix}
\begin{pmatrix} \phi_{2x}\\\phi_{2y}\end{pmatrix}.
\end{equation}
By considering the eigenvectors in Eqs.~(\ref{linear unbroken}) or~(\ref{nonlinear unbroken solution}), and by letting $v_1=\begin{pmatrix} 1&e^{i\theta}\end{pmatrix}^{T}$ and
$v_2=\begin{pmatrix} 1&-e^{-i\theta}\end{pmatrix}^{T}$, any initial state $v_0=\begin{pmatrix} a_{01}&a_{02}\end{pmatrix}^{T}$ can then be projected. Absolute values of the complex coefficients $c_1 = \langle v_0 | v_1\rangle$ and $c_2 = \langle v_0 | v_2\rangle$ associated with these eigenvectors can then be obtained and are given by,
\begin{multline}
|c_1|^2 = |a_{01}|^2+|a_{02}|^2 \\+2|a_{01}||a_{02}|\cos(\Delta\phi_0 -\theta),\end{multline}
\begin{multline}
|c_2|^2 = |a_{01}|^2+|a_{02}|^2 \\-2|a_{01}||a_{02}|\cos (\Delta\phi_0 +\theta),
\end{multline}
where $\Delta\phi_0 = \phi_{a_{01}}-\phi_{a_{02}}$ is the initial phase difference between $a_{01}$ and $a_{02}$. As we will see, the eigenvector with the larger initial amplitude will eventually dominate.

The eigenvalue for the vector $v_1$ is $\lambda_1 = \cos\theta$ which implies a counter-clockwise rotation in the complex plane. Note that this corresponds to the low frequency supermode since the fast variations in the field, leading to Eq.~(\ref{model}) were assumed to be of the form $e^{-i\omega_0 t}$ where $\omega_0$ is the resonance frequency of each individual resonator. Similarly the other eigenvalue, $\lambda_2 = -\cos\theta$ corresponds to its high frequency counterpart.

For counter-clockwise rotation, one requires that $|c_1|^2>|c_2|^2$ and vice versa for clockwise rotation. Since $\cos\theta>0$ for $\theta=\left[0,\pi/2\right]$, these conditions can be reduced to, (i)$\Delta\phi_0\leq\pi/2$, where the low frequency supermode survives and (ii)$\Delta\phi_0>\pi/2$, favoring its high frequency counterpart.
As an example, let $g_0=2.25$, $f_0=1$ and $\gamma=0.1$, that satisfy the conditions in Eq.~(\ref{gain range NL unbroken}). We now consider two cases for the phase difference in the initial values $a_{01}$ and $a_{02}$. Figure ~\ref{fig6}(b) shows the field evolution when the initial amplitudes in the two rings are equal, $|a_{01}|=|a_{02}|=0.2$, but the phase difference is $\Delta\phi_0=\pi/2+0.1$ and (c) shows the same case when $\Delta\phi_0=\pi/2-0.1$. The intensity evolution with time for these two cases happens to be identical and is depicted in Fig.~\ref{fig6}(a).
The initial exponential growth is evident and the intensities finally saturate to a common value as given by Eq.~(\ref{nonlinear unbroken solution}). We note that in an actual experiment, both eigenmodes will be excited from noise and hence the spectrum will involve two lines at $\pm \kappa\cos(\theta_\text{nl})$ around $\omega_0$.
\section{\label{S5}General system behavior}
Based on the results of the previous section, it can be established that in the steady state, the form of the nonlinear solutions is predetermined by the system parameters, specifically by the normalized values of unsaturated absorption ($f_0$) and linear loss ($\gamma$). In the coupled ring resonator arrangement, as the pumping level is increased (as $g_0$ increases), there are two possible scenarios for the system behavior. If $(\gamma+f_0)<1$, lasing begins in the linear unbroken PT-symmetric domain (Eq. (\ref{linear unbroken})) and then moves into the nonlinear unbroken PT-symmetric regime where the field intensities are equal in both rings albeit with a phase difference, according to Eq.~(\ref{nonlinear unbroken solution}). If on the other hand $(\gamma+f_0)>1$, lasing starts in the linear broken PT-symmetric domain (Eq.~(\ref{linear broken})) and then transitions into the nonlinear broken PT-symmetric phase where the distribution of field strengths in the two coupled resonators is asymmetric and a phase difference of $\pi/2$ exists between them, as established in Sec.~\ref{S4-1}. At even higher gain levels, interestingly, a phase transition occurs from the broken domain into the nonlinear unbroken PT domain when $g_0>g_c$.
\begin{figure}
  \includegraphics{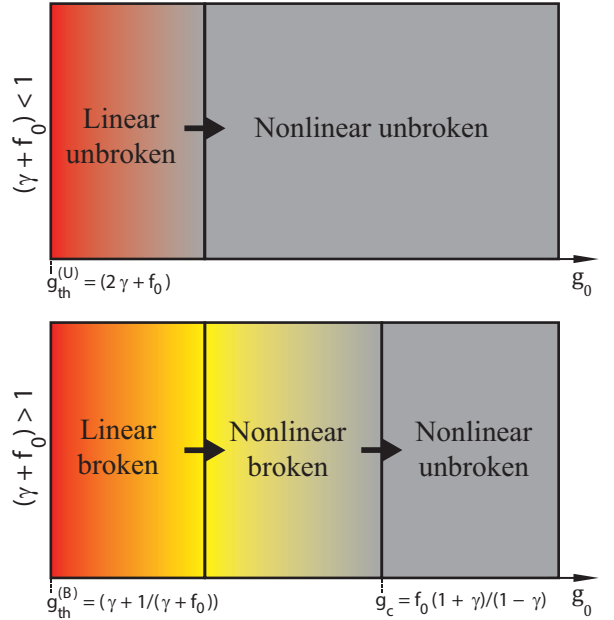}
  \centering
  \caption{\label{fig7}System response as a function of gain in two different parameter regimes is schematically shown. In the upper half, where $(\gamma+f_0)<1$, the system is always in an unbroken PT phase. In the lower half, however, where $(\gamma+f_0)>1$, the configuration first transitions from a linear broken to a nonlinear broken phase and then eventually enters the nonlinear unbroken domain when $g_0$ exceeds $g_c$.}
\end{figure}
The two scenarios are summarized in Fig.~\ref{fig7}, where the nonlinear reversal of a PT-symmetric phase transition (broken to unbroken) is displayed in the lower half. It should be noted that the lasing thresholds in these two cases are different and both paths eventually end up in the unbroken PT-like phase as the gain is increased.
\section{\label{S6}Experimental results}
\begin{figure}
  \includegraphics{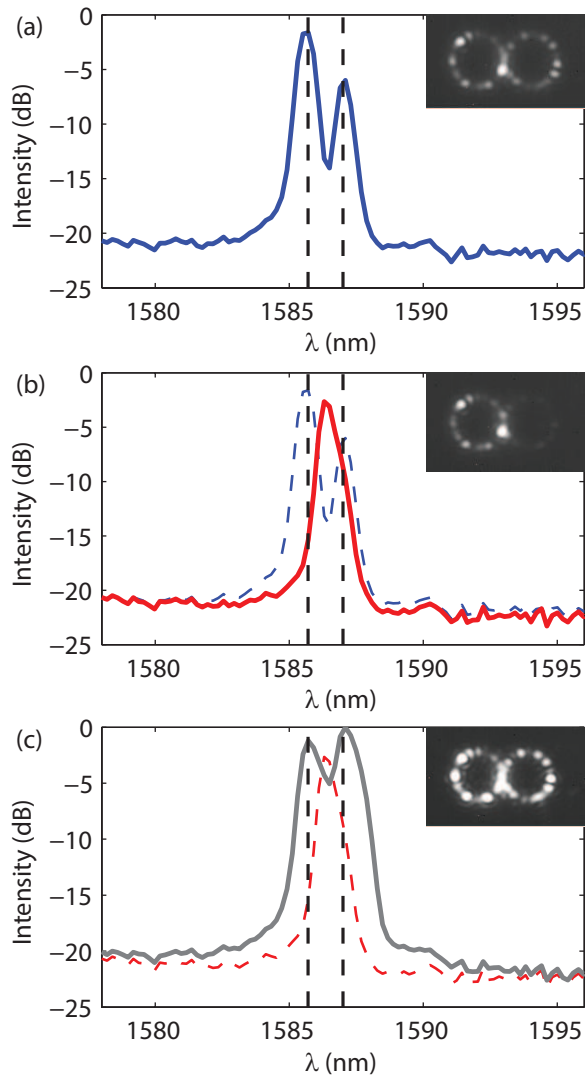}
  \centering
  \caption{\label{fig8}Emitted spectrum from (a) uniformly pumped coupled microrings with the pump power of 0.4 mW (b) PT-symmetric structure when 0.4 mW of pump power reaches the active ring (c) PT-symmetric structure when the active ring is subjected to 0.8 mW pump power. The insets depict mode profiles of the different scenarios, recorded by the scattering from the surface of the rings. Dashed vertical lines are used to compare the locations of resonances.}
\end{figure}

To experimentally verify our findings, we used lithographic techniques to fabricate sets of coupled microring resonators comprised of six InGaAsP (Indium-Gallium-Arsenide-Phosphide) quantum wells embedded in InP, capable of providing amplification in the wavelength range $1350\text{-}1600$ nm. A detailed description of the fabrication process can be found in Ref.~\cite{Hodaei}. The rings in our experiments have an outer radius of $10$ $\mu$m, a width of $500$ nm, and a height of $210$ nm. Such dimensions are deliberately chosen so as the rings support a single transverse mode and to also favor the TE polarization. At first, the two coupled resonators were evenly illuminated using a circular pump beam with a diameter of $80$ $\mu$m. The intensity distribution and spectrum of the modes in the microrings are monitored using a CCD camera and a spectrometer respectively. Figure~\ref{fig8}(a) shows the spectrum of the two active rings when are both exposed to a peak pump power of $0.4$ mW (15 ns pulses with a repetition rate of 290 kHz).
Under these conditions, coupling-induced mode splitting can clearly be seen. Next, a knife edge is used to selectively withhold the pump power from one of the rings, hence establishing a PT-symmetric gain/loss microring arrangement. Figure~\ref{fig8}(b) illustrates the lasing spectrum of this PT system. As expected, lasing occurs exclusively in the active cavity and the frequency of the resonance shifts to the center of the supermode peaks. In other words, the system starts lasing in the broken PT-phase as predicted in Sec.~\ref{S4-1}. Next, the pump power illuminating the active ring is increased by a factor of two while keeping again the lossy ring in the dark. The emission spectrum of the PT arrangement subjected to such a high pump power is depicted in Fig.~\ref{fig8}(c). In agreement with our theoretical predictions (Sec.~\ref{S4-2}), the PT-symmetry of the combined structure is now restored due to a saturation of nonlinearities. In this regime, both resonators are again contributing equally to lasing and as a result two supermode wavelength peaks are now present in the measured spectrum. Our experimental results confirm the fact that nonlinear processes are indeed capable of reversing the order in which the symmetry breaking occurs.

The discussions in earlier sections are also applicable to the findings in Ref.~\cite{Hodaei}. In that work, lasing was observed when both microrings were at first equally illuminated (in a way similar to Fig.~\ref{fig8}(a)), in which case the system was positioned in the unbroken PT-symmetry phase. This behavior is in agreement with our theoretical results presented in Sec.~\ref{S4-2} provided that one sets $f_0=-g_0$. In this case, $\eta$ is purely imaginary and $\theta_{\text{nl}}=0$, and hence the normalized eigenvalues are $\lambda_{1,2}=\pm 1$, i.e. the mode splitting is twice the coupling between the two cavities\textemdash resembling that in standard Hermitian systems. On the other hand, by removing the pump from one of the rings, saturable losses are introduced since now $f_0$ is positive. In this scenario, Eq.~(\ref{gain range NL unbroken}) is no longer satisfied and as a result the system enters the broken PT phase, in agreement with the observed behavior in Ref.~\cite{Hodaei}.
\section{\label{S6}Conclusions}
In conclusion, we have shown that a nonlinear dimer of two coupled microring laser resonators can transition from a nonlinear broken PT-symmetric phase into an unbroken PT domain provided that lasing was initiated in a broken mode. This surprising result is a byproduct of nonlinear saturation that is capable of re-establishing the PT-symmetric phase, something that is not possible in the linear domain. On the other hand, if lasing initially occurs in an unbroken mode, the system always remains in the unbroken phase even under nonlinear conditions. In all cases, in spite of the presence of nonlinearities, the eigenmodes of this arrangement retain features associated with linear eigenvectors in PT-symmetric configurations.
\begin{acknowledgements}
The authors gratefully acknowledge the financial support from NSF CAREER Award (ECCS-1454531), NSF (grant ECCS-1128520), and AFOSR (grants FA9550-12-1-0148 and FA9550-14-1-0037), ARO (grant W911NF-14-1-0543).
\end{acknowledgements}

\bibliography{REFS}

\end{document}